\pdfoutput=1
\documentclass[aps,prl,twocolumn,showpacs,groupedaddress]{revtex4}  

\usepackage{graphicx}
\usepackage{dcolumn}
\usepackage{bm}
\usepackage{amssymb}   

\def\MET{{\mbox{$E\kern-0.57em\raise0.19ex\hbox{/}_{T}~$}}}
\def\METnoSpace{{\mbox{$E\kern-0.57em\raise0.19ex\hbox{/}_{T}$}}}

\begin{document}


\hspace{5.2in} \mbox{FERMILAB-PUB-09-229-E}

\title{Search for dark photons from supersymmetric hidden valleys}
%
\author{V.M.~Abazov$^{37}$}
\author{B.~Abbott$^{75}$}
\author{M.~Abolins$^{65}$}
\author{B.S.~Acharya$^{30}$}
\author{M.~Adams$^{51}$}
\author{T.~Adams$^{49}$}
\author{E.~Aguilo$^{6}$}
\author{M.~Ahsan$^{59}$}
\author{G.D.~Alexeev$^{37}$}
\author{G.~Alkhazov$^{41}$}
% ...

\affiliation{\vspace{0.1 in}(The D\O\ Collaboration)\vspace{0.1 in}}
% ...
\affiliation{$^{6}$University of Alberta, Edmonton, Alberta, Canada;
                Simon Fraser University, Burnaby, British Columbia, Canada;
                York University, Toronto, Ontario, Canada and
                McGill University, Montreal, Quebec, Canada}
% ...
\affiliation{$^{30}$Tata Institute of Fundamental Research, Mumbai, India}
% ...
\affiliation{$^{37}$Joint Institute for Nuclear Research, Dubna, Russia}
% ...
\affiliation{$^{41}$Petersburg Nuclear Physics Institute,
                St. Petersburg, Russia}
% ...
\affiliation{$^{49}$Florida State University, Tallahassee, Florida 32306, USA}
% ...
\affiliation{$^{51}$University of Illinois at Chicago,
                Chicago, Illinois 60607, USA}
% ...
\affiliation{$^{59}$Kansas State University, Manhattan, Kansas 66506, USA}
% ...
\affiliation{$^{65}$Michigan State University,
                East Lansing, Michigan 48824, USA}
% ...
\affiliation{$^{75}$University of Oklahoma, Norman, Oklahoma 73019, USA}
% ...
\date{May 11, 2009}

\begin{abstract}
We search for a new light gauge boson, a dark photon, with the D0 experiment. 
In the model we consider, supersymmetric partners are pair produced and cascade to lightest neutralinos 
that can decay into the hidden sector state plus either a photon or a dark photon. The dark photon decays through 
its mixing with a  photon into fermion pairs. We therefore investigate a previously unexplored final state
that contains a photon, two spatially close leptons, and large missing transverse energy.
We do not observe any evidence for dark photons and set a limit on their production.
\end{abstract}

\maketitle


Hidden valley models \cite{hv} introduce a new hidden sector, which is very weakly coupled to the 
standard model (SM) particles, and therefore can easily escape detection. An important 
subset of hidden valley models also contain supersymmetry (SUSY), a fundamental symmetry between 
fermions and bosons postulating the existence of SUSY partners. 
At colliders, in the case of $R$-parity conservation~\cite{rpar}, 
superpartners are produced in pairs and decay to the SM particles and the lightest superpartner (LSP). 
The LSP is a stable, weakly interacting particle,  and can not be detected in collider detectors.

Recently, these models were called upon to explain the results of several cosmic ray 
detection experiments~\cite{PAMELA, ATIC}. Taken together with other experiments,
including new results from Fermi/LAT \cite{fermi},
there is evidence of an excess of high energy positrons and no excessive production of anti-protons or photons.
The  excess can be attributed ~\cite{nima} to the dark matter particles annihilating into
pairs of new light gauge bosons, dark photons, which are force carriers in the hidden sector.
The dark photon mass can not be much larger than 1 GeV
to give rise to Sommerfeld enhancement \cite{somm} of the dark matter annihilation cross section, and not to decay 
into neutral pions and/or baryons. The masses of the hidden sector states are also around 1 GeV,
with mass splitting around MeV, thus providing a possible explanation of
the DAMA \cite{DAMA} signal through "inelastic Dark Matter" scenarios.
Dark photons decay through mixing with photons into SM fermions with branching fractions that 
can be calculated from the measurements~\cite{pdg} of  
$R = \sigma(e^+e^- \rightarrow {\rm hadrons})/\sigma(e^+e^- \rightarrow \mu^+\mu^-)$, 
and strongly depend on the dark photon mass. For dark photon masses ($m_{\gamma_D}$) 
below the dimuon threshold of $\simeq 200$~ MeV, only decays into electrons are possible.
For $m_{\gamma_D} \simeq 0.5$~ GeV the decay rates into electrons and muons are 
approximately 40\% each. The lowest value
of the leptonic branching (3.7\%) occurs if the dark photon mass is accidentally equal to that of the $\phi$ meson.

\begin{figure}
\begin{center}
\includegraphics[width=8cm]{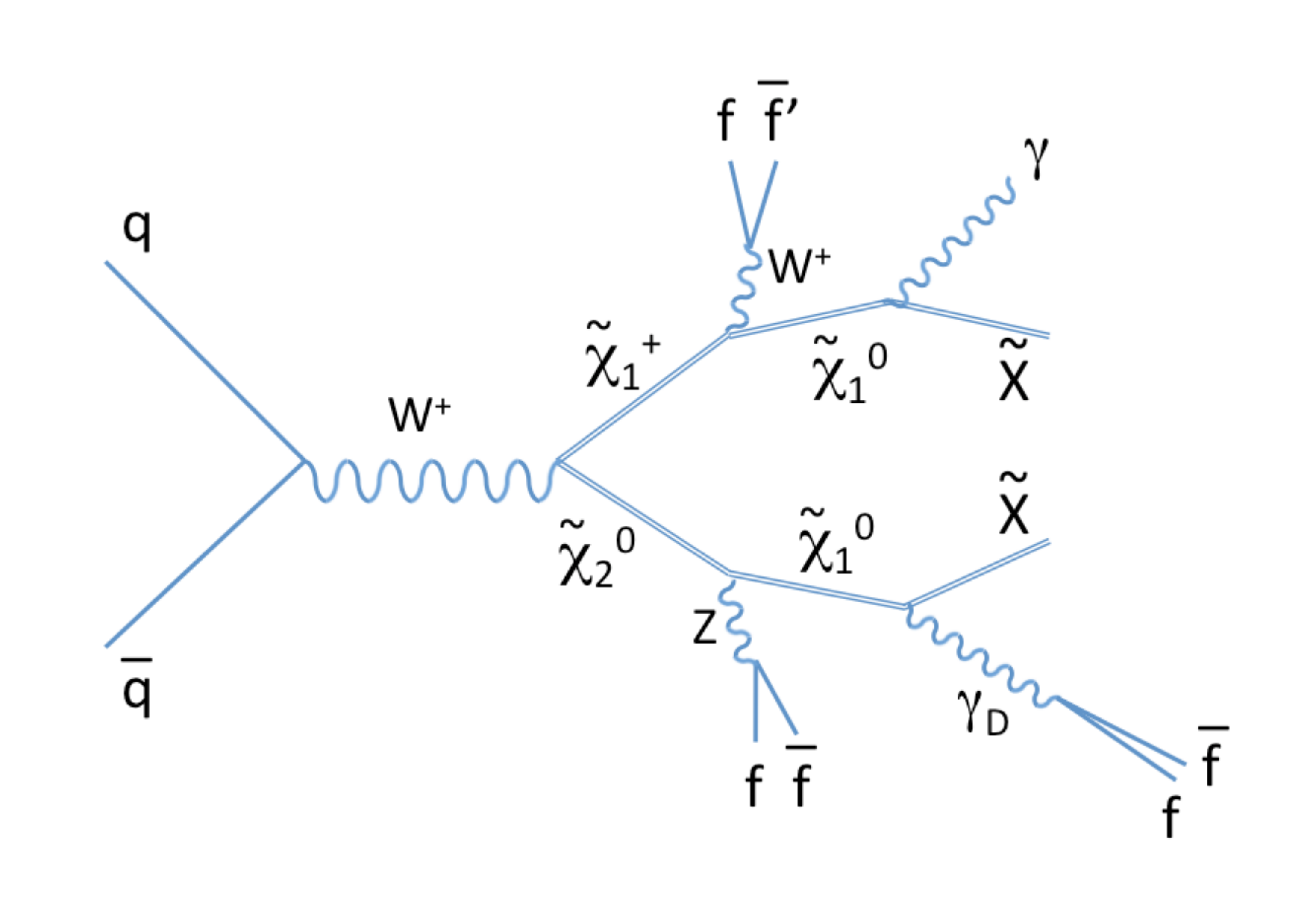}
\end{center}
\caption{\label{fig:diagram1} One of the diagrams giving rise to the events with a photon, dark photon ($\gamma_D$), 
and large missing energy due to escaping darkinos ($\tilde{X}$) at the Fermilab Tevatron Collider. }
\end{figure}

In this Letter we will follow the phenomenological scenario developed in~\cite{dg-thomas}. 
A diagram of a possible process at the Fermilab Tevatron Collider is shown in Figure~\ref{fig:diagram1}. 
Gauginos are pair produced and decay into SM particles 
and the lightest neutral gaugino (neutralino, $\tilde{\chi}^0_1$), which in turn decays with comparable branching ratios
into either a hidden sector darkino $\tilde{X}$ 
(which is the LSP), and a photon, or into darkino and a dark photon ($\gamma_D$). 
Hadronic dark photon decays are overwhelmed by SM jet backgrounds. Thus, we only consider 
dark photon decays into isolated electron or muon pairs. 
Both darkinos escape detection and result in large missing transverse energy (\METnoSpace). 
The branching fraction of the neutralino into the dark photon, 
${\cal B} = Br(\tilde{\chi}^0_1 \rightarrow \gamma_D \tilde{X})$,
is a free parameter of the model. If it is small, the decays into a  photon dominate, and 
signature is the same as of 
GMSB SUSY~\cite{gmsb_susy} with the neutralino as next-to-lightest superpartner (NLSP). 
Larger values of ${\cal B}$ give rise to events 
where one of the two neutralinos decays into a dark photon, resulting in a final state with one photon, 
two spatially close (and therefore not satisfying traditional isolation requirements) leptons and large \METnoSpace.

This Letter describes a search for this, so far unexplored, final state in $p\bar{p}$ collisions at a center
of mass energy of 1.96 TeV recorded by the D0 detector~\cite{d0detector} at the Fermilab Tevatron Collider.
As is described below, our search is optimized for low dark photon masses,  $m_{\gamma_D} < 2.5$ GeV.
Note that it is also sensitive to the case where the neutralino decays into a hidden state
$\tilde{Y}$ with somewhat higher mass than the dark photon. 
The $\tilde{Y}$ may cascade down to the darkino through other hidden states which may be
long-lived and can result in the emission of highly collimated low energy SM particles, some of which
could be leptons. Most of the energy of the $\tilde{Y}$ will stay in the hidden sector 
and therefore the high \MET should not be substantially reduced.  This analysis is also sensitive to another 
possible scenario, proposed in~\cite{thaler}, in which a light axion that decays into muon pairs takes the place 
of the dark photon in the decays described above.

Data for this analysis correspond to an integrated luminosity of 4.1 fb$^{-1}$ after application
of data quality and trigger requirements. Events must satisfy a set of high transverse energy ($E_T$), single
electromagnetic (EM) cluster triggers which are fully efficient for photons with
$E_T> 30$~ GeV.

EM clusters are selected from calorimeter clusters, built using the simple 
cone algorithm of radius ${\cal R} = \sqrt{(\Delta\eta)^{2} + (\Delta\phi)^{2}} = 0.4$  ~\cite{d0coord},
by requiring that the fraction of the energy deposited in the EM section
of the calorimeter, ${\rm EM}_{frac}$, is above 95\% and the calorimeter isolation variable
${\cal I} = [E_{tot}(0.4) - E_{EM}(0.2)]/E_{EM}(0.2)$ is less than 0.2, 
where $E_{tot}(0.4)$ is the total energy in a cone of radius ${\cal R}=0.4$, 
corrected for the underlying event contribution, and
$E_{EM}(0.2)$ is the EM energy in a cone of radius ${\cal R}=0.2$, which is taken to be
the EM cluster energy.

Photon candidates are selected from central calorimeter ($|\eta|<1.1$) EM clusters which have 
(i) ${\rm EM}_{frac} > 97\%$, (ii) ${\cal I}<0.07$, (iii) a
shower shape consistent with that of a photon, and  (iv) the scalar
sum of the transverse momenta ($p_T$) of all tracks originating
from the primary vertex in an annulus $0.05 < {\cal R} < 0.4$
around the cluster less than 2 GeV.  
Additionally, we require that photon candidates are not spatially matched to
activity in the tracker. The tracker activity can either be a reconstructed charged particle's track or
a density of hits in the silicon microstrip and central fiber trackers consistent with a track.
The EM clusters that do not have matched activity in the tracker, but fail other photon selection criteria, 
are dominated by jets that have fragmented into neutral pions, and are referred to below as fake photons.

We search for dark photons in events with at least one photon with $E_T > 30$ GeV and $\MET > 20$ GeV
($\MET$ is computed using all calorimeter cells and corrected for EM and jet energy scales).
Dark photon candidates are formed by selecting pairs of oppositely charged spatially close (${\cal R} < 0.2$) tracks 
that originate from the same point ($|\Delta z |< 2$ cm) along the beam line. 
The leading (trailing) track $p_T$ is required to exceed 10 (5) GeV.
We then require the scalar sum of $p_T$ of all tracks excluding the pair in a cone of radius 0.4 centered on 
the pair momentum direction to be less than 2 GeV. To reduce the QCD background we 
require that each track must have its azimuthal angle not aligned with a photon, 
$0.4 < \Delta\phi_{\gamma, track} < 2.74$.
In rare cases, when there is more than one such pair in the event, we select 
the one with the highest trailing track $p_T$. 

For a dark photon decaying into a pair of electrons, the calorimeter depositions 
overlap, so we require that the dark photon candidate matches an EM
cluster with $E_T > 10$ GeV, ${\rm EM}_{frac} > $ 97\%, and ${\cal I} < 0.1$.
For the dimuon decay mode, we require that at least one of the tracks is matched to a reconstructed muon, and the
energy deposited in the calorimeter in the annulus $0.1 < {\cal R} < 0.4$  is below 3 GeV.

Dark photons would manifest themselves as a narrow peak in the lepton pair invariant mass distribution.
We use a Monte Carlo simulation to characterize the mass resolution, as well as the efficiency to reconstruct the events.
{\sc susyhit}~\cite{susy-hit} is used to calculate masses and decay 
probabilities for the GMSB SUSY ~\cite{snowmassE} model, known as Snowmass Slope SPS 8,
and produce the Les Houches Accord~\cite{lha} card files. These files are modified to introduce
neutralino decays to a dark photon. Events with one of the two neutralinos
decaying into a dark photon and the other decaying into a photon are generated with {\sc pythia}~\cite{pythia}
using {\sc cteq6l1} parton distributions~\cite{cteq} and are
passed through the full {\sc geant}-based~\cite{geant} detector simulation and the same reconstruction chain
as the data.  Following~\cite{gmsb}, the leading order (LO) signal cross sections calculated by {\sc pythia}
are scaled to match the next-to-leading order (NLO) prediction using $k$-factor values extracted from~\cite{kfactor}.
The event kinematics depends on the mass of the dark photon and the masses of superpartners,
resulting in slight variations in mass resolution and selection efficiency. Typical values are 5\% and 12\%, respectively.

There are three types of SM processes that contribute to our data sample:
\begin{enumerate}
\item[{\bf B1}] QCD events with real or fake photons and mismeasured \METnoSpace. These contain jets or photon 
conversions faking the dark photon.
\item[{\bf B2}]  $W \rightarrow e/\mu ~ \nu $ plus a real or fake photon. The dark photon is faked by a accidental 
overlap of a high $p_T$ track with the lepton.
\item[{\bf B3}] $W \rightarrow  \tau\nu \rightarrow 3h^\pm \nu$ plus real or fake photon. One of the particles from 
$\tau$ lepton decays is lost or very soft, and the remaining $\tau$ decay products fake the dark photon.
\end{enumerate}

We study dark photon candidate mass distributions in three control samples where we do not 
expect dark photons to appear.  The $QCD\_\gamma$ control sample
is selected by reversing the \MET cut. The $QCD\_{jet}$ sample is
selected by using the same criteria as the $QCD\_\gamma$ sample, but requiring a fake
photon instead of a photon. Finally, the  $QCD\_W$ sample requires a fake photon and $\MET > 20$~ GeV.  
All three have contributions from {\bf B1}, although the relative fraction of multijets, single photon production, and
diphoton production varies  among the three control samples. Backgrounds {\bf B2} and {\bf B3}, however, can only 
significantly contribute to the $QCD\_W$ sample. We observe no difference between the dark photon candidate 
mass distributions in the three control samples. We therefore conclude that the background to dark photon production is dominated 
by {\bf B1}, and use the average shape of the dark photon candidates mass distributions in all control samples as our background model.

\begin{figure}
\includegraphics[width=8.5cm]{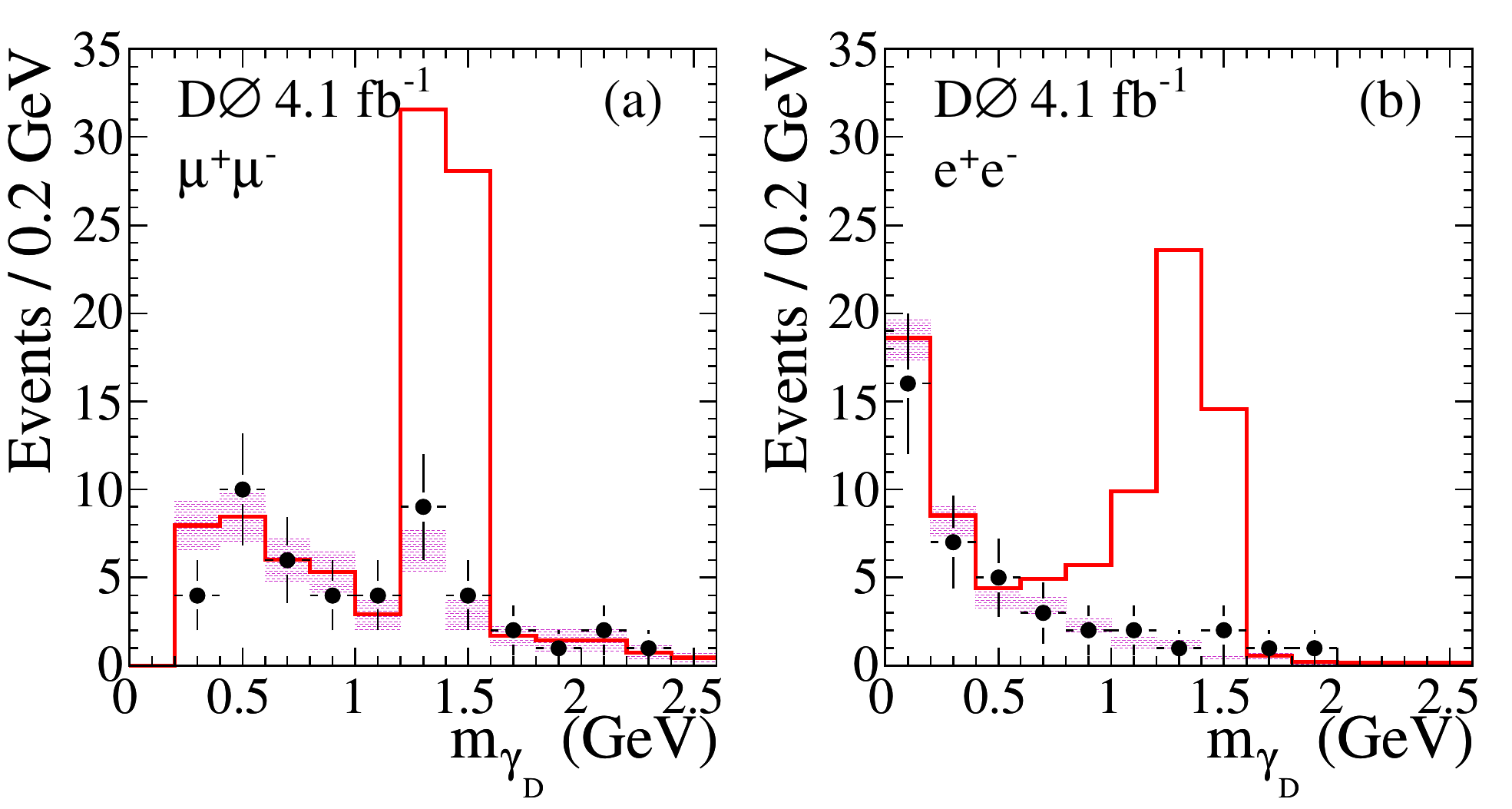}
\caption{\label{fig:mlimit} Observed mass distributions in the signal region are represented as points with error bars, the background estimation is shown as filled band, and an example signal for m$_{\gamma_D} = 1.4$~ GeV plus background is shown as the solid histogram for the dimuon channel (a) and the dielectron channel (b).}
\end{figure}

The dark photon candidate invariant mass distributions in the signal sample are shown in Figure~\ref{fig:mlimit} separately for the electron and muon channels, together with the expected contribution from dark photons with 
a mass of 1.4 GeV.

We see no evidence of a dark photon signal and proceed to set limits on its production. To set limits we use the standard D0~likelihood fitter~\cite{collie} that incorporates a log-likelihood ratio (LLR) statistic method \cite{cls}. The value of $CL_s$ is defined as $CL_s = CL_{s+b}/CL_{b}$, where $CL_{s+b}$ and $CL_{b}$
are the confidence levels for the signal plus background hypothesis and the background-only (null) hypothesis, respectively. These confidence levels are evaluated by integrating the corresponding $LLR$ distribution populated by simulating outcomes via Poisson statistics. Systematic uncertainties are treated as uncertainties on the expected
number of signal and background events, not the outcomes of the limit calculations. This approach ensures that the
uncertainties and their correlations are propagated to the outcome with their proper weights. 
The limit is set by simultaneously fitting dilepton invariant mass 
distributions in data for the muon and electron channels to the signal
and background predictions for each signal point, defined by the dark photon and the lightest chargino masses.
For each dark photon mass the background is normalized outside of the expected signal region.
The systematic uncertainty on the signal reconstruction efficiency (25\%) is dominated by the uncertainty to 
reconstruct the two spatially close tracks from the dark photon decays (20\%). The latter was cross-checked 
with data using tau decays and converted photons in radiative $Z\rightarrow \mu\mu\gamma$ decays.
We also took into account the uncertainty on the total integrated luminosity (6.1\%) and 
the effect of varying the dark photon mass resolution by 10\%.

\begin{figure}
\includegraphics[width=8.5cm]{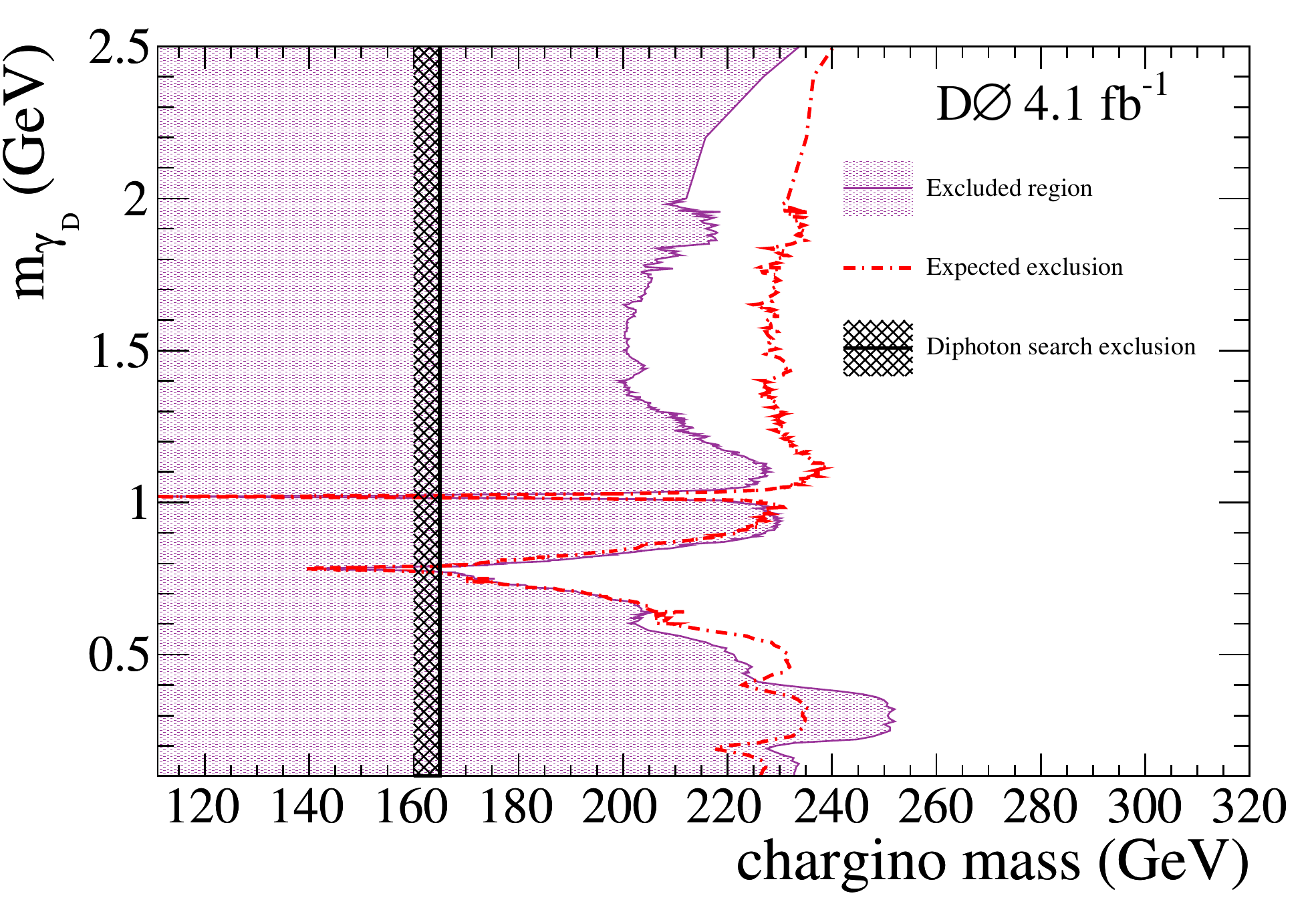}
\caption{\label{fig:exclusion} The excluded region of possible masses of the lightest chargino and the dark photon
for ${\cal B}=0.5$ are shown as the shaded region. The expected limit is illustrated as the dash-dotted line. 
The vertical black line corresponds to the exclusion from the diphoton search~\cite{gmsb}. }
\end{figure}

\begin{figure}
\includegraphics[width=8.5cm]{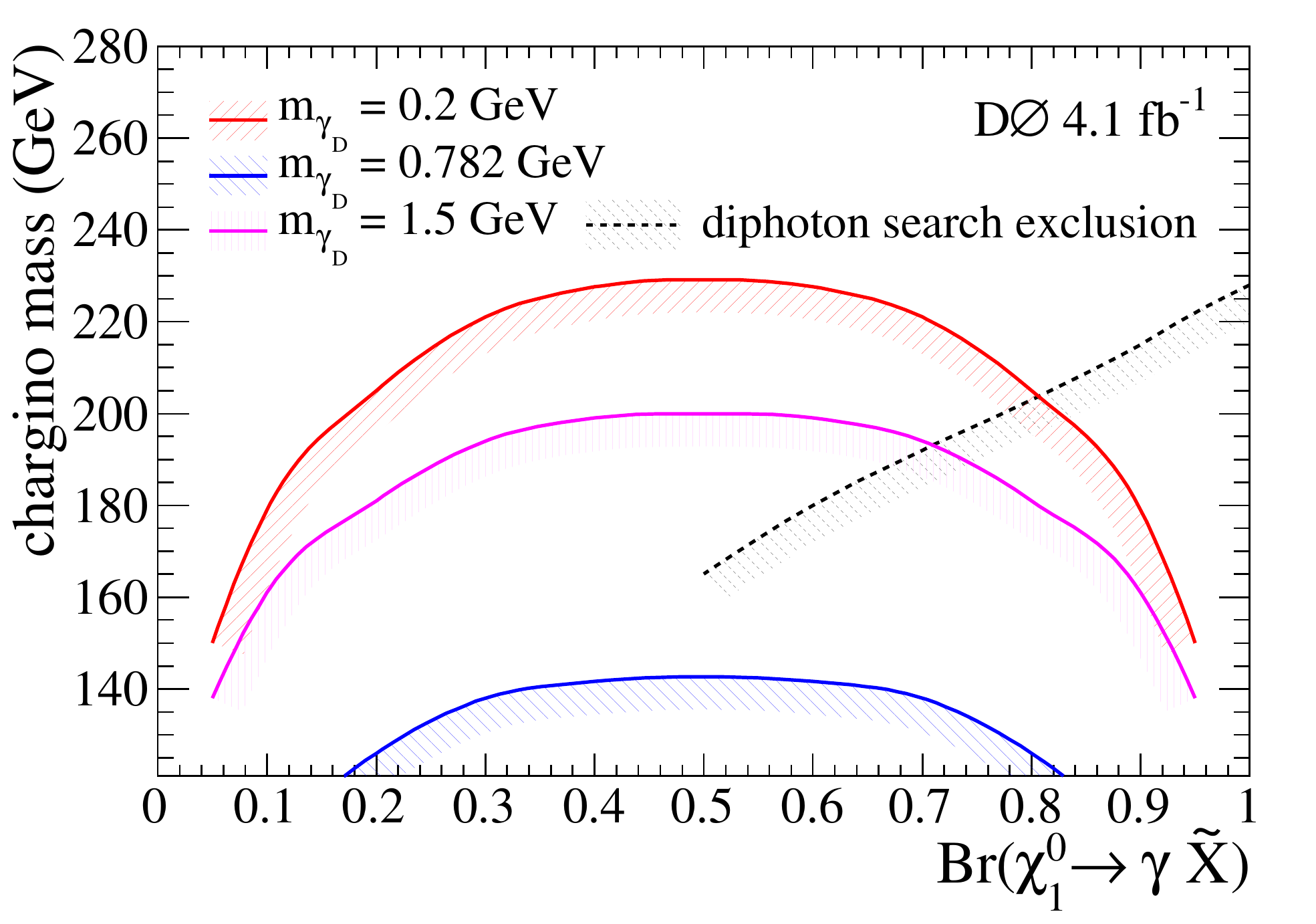}
\caption{\label{fig:exclusion-br}  The dependence of the excluded chargino masses on the branching ratio of the neutralino into a  photon are given for dark photon masses of 0.2, 0.782, and 1.5 GeV.  The black contour
corresponds to the exclusion from the diphoton search~\cite{gmsb}. }
\end{figure}

We interpret the cross section limits as limits on the lightest chargino mass as a function of the dark photon mass and 
the neutralino branching fraction. For ${\cal B} = 0.5$
the excluded region of chargino and dark photon masses is shown in Figure~\ref{fig:exclusion}.
In Figure~\ref{fig:exclusion-br} we display the chargino mass limit as a function of ${\cal B}$ for
three representative dark photon masses: 0.2 GeV (only the electron channel is open), 
0.782 GeV (low branching fraction into leptons due to $\omega$ and $\rho$ mesons), and 1.5 GeV. 
Our previous limit on the GMSB SUSY in the diphoton final state~\cite{gmsb} is directly applicable 
to the model considered in this Letter, although it does not probe the dark photon mass. 
The corresponding exclusion contours are shown in Figures~\ref{fig:exclusion}~and~\ref{fig:exclusion-br}.

To summarize, we search for a previously unexplored final state consisting of a photon, two spatially close leptons from 
hypothetical dark photon decays and large missing energy. We find no evidence for such events, and set limits on their production in a benchmark model~\cite{dg-thomas}.
For dark photon masses of 0.2, 0.782, and 1.5 GeV we exclude chargino masses of 230, 142, and 200 GeV, respectively.

We would like to thank Scott Thomas and David Shih for many inspiring discussions and help with the signal simulation. 
%
We thank the staffs at Fermilab and collaborating institutions, 
and acknowledge support from the 
DOE and NSF (USA);
CEA and CNRS/IN2P3 (France);
FASI, Rosatom and RFBR (Russia);
CNPq, FAPERJ, FAPESP and FUNDUNESP (Brazil);
DAE and DST (India);
Colciencias (Colombia);
CONACyT (Mexico);
KRF and KOSEF (Korea);
CONICET and UBACyT (Argentina);
FOM (The Netherlands);
STFC and the Royal Society (United Kingdom);
MSMT and GACR (Czech Republic);
CRC Program, CFI, NSERC and WestGrid Project (Canada);
BMBF and DFG (Germany);
SFI (Ireland);
The Swedish Research Council (Sweden);
CAS and CNSF (China);
and the
Alexander von Humboldt Foundation (Germany).
%


\begin{thebibliography}{99}
\bibitem{d0coord}The D0 detector utilizes a right-handed coordinate system with the $z$-axis 
pointing in the direction of the proton beam and the $y$-axis pointing
upwards. 
The azimuthal angle $\phi$ is defined in the $xy$ plane
measured from the $x$-axis. 
The pseudorapidity is defined as $\eta = -\ln[\tan(\theta/2)]$,
where $\theta = \arctan(\sqrt{x^2+y^2}/z)$.
\end{thebibliography}
\end{document}